\documentclass{article}
\usepackage{graphicx}
\usepackage[english]{babel}
\usepackage{srcltx}
\usepackage{amsmath}
\usepackage{amssymb}
\usepackage{amsfonts}
\usepackage{latexsym}
\usepackage{textcomp}
\usepackage{appendix}
\usepackage{multirow}
\usepackage{booktabs}
\usepackage{rotating}
\usepackage{url}
\usepackage{array}
\usepackage{marvosym}
\usepackage{afterpage}
\usepackage{pdflscape}
\usepackage[table,svgnames]{xcolor}
\usepackage{graphics}
\usepackage{graphicx}
\usepackage{caption}
\usepackage{multirow}
\usepackage{booktabs}
\usepackage{standalone}
\usepackage{tabulary}
\usepackage{adjustbox}
\usepackage{booktabs}
\usepackage{pdflscape,afterpage,caption}
\usepackage{longtable}
\usepackage[round,authoryear]{natbib}
\usepackage{subfig} 
\graphicspath{{./plots/}}
\usepackage{lineno,hyperref}
\modulolinenumbers[5]
\usepackage[margin=1.25in]{geometry}

\title{Where do we stand in cryptocurrencies economic research? A survey based on hybrid analysis}

\author{Aurelio F. Bariviera\thanks{Corresponding author: aurelio.fernandez@urv.cat}, Ignasi Merediz-Sol\`a  \\
 \scriptsize Universitat Rovira i Virgili, Department of Business, Av. Universitat 1, 43204 Reus, Spain}

\begin{document}
\maketitle

\begin{abstract}
This survey develops a dual analysis, consisting, first, in a bibliometric examination and, second, in a close literature review of all the scientific production around cryptocurrencies conducted in economics so far. The aim of this paper is twofold. On the one hand, proposes a methodological hybrid approach to perform comprehensive literature reviews. On the other hand, we provide an updated state of the art in cryptocurrency economic literature. Our methodology emerges as relevant when the topic comprises a large number of papers, that make unrealistic to perform a detailed reading of all the papers. This dual perspective offers a full landscape of cryptocurrency economic research. Firstly, by means of the distant reading provided by machine learning bibliometric techniques, we are able to identify main topics, journals, key authors, and other macro aggregates. Secondly, based on the information provided by the previous stage, the traditional literature review provides a closer look at methodologies, data sources and other details of the papers. In this way, we offer a classification and analysis of the mounting research produced in a relative short time span.
\end{abstract}

\section{Introduction \label{sec:intro}} 

Cryptocurrency literature has been experimenting a sustained growth. As a new object of study, cryptocurrencies offer a rich field to implement both old and new methodologies, in order to uncover the salient characteristics of this market. After some years of continuous research, it is necessary to draw a situation map of current research and comment of literature gaps and research perspectives. In this sense this work precisely aims at becoming a reference guide for researchers. We developed our paper in two complementary steps. First, we implement a biblometric analysis, in order to get the most relevant features arising from text mining analysis of titles, abstracts, keywords, authors and journal titles. Second, we produce an in-depth analysis of 98 papers, from the most important journals detected in the previous step. 

There are some previous experiences of literature review, but with a broader scope. \cite{Liu2016} uses exclusively co-word analysis of 256 papers from Scopus database, in order to classify them into technological, economic and legal aspects of bitcoin. \cite{Miau2018} and \cite{Holub2018} analyze the whole blockchain research area.

The two closest papers to ours are \cite{Corbet2019} and \cite{MeredizBariviera2019}. The first one produces a systematic review of fifty-two quantitative investigations of cryptocurrency markets. The second one, provides a classification and identification of key elements of 1162 papers dealing with bitcoin, across different disciplines. Our methodological approach is different. To the best of our knowledge, this is the first paper that  combine bibliometric analysis and close literature review into the same paper, in order to produce a comprehensive landscape of the current cryptocurrency research exclusively within economics.

On the one hand, bibliometric analysis provides a semi-authomatic classification of papers, using machine learning. This first approach is very useful, specially when considering an large number of papers. On the other hand, in-depth reading of individual papers helps to identify methodologies, datasets, and results. As a consequence, this paper harmonizes machine-based classification with the insight of the specialized reader. 

Our paper contributes to the literature in several ways: (i) it presents a hybrid methodology, by combining distant (bibliometric) and close (in-depth) reading in order to produce a literature survey; (ii) it comprises more up-to-date literature by considering also articles in press, in addition to those already abstracted in Scopus or Web of Science; (iii) it allows to infer emerging research lines in cryptocurrency literature. 

The rest of the paper is structured as follows.  Section \ref{sec:bibliometric}  describes the data set and comments the main findings of our bibliometric study. Based on these results, Section \ref{sec:literature} works with a new dataset and produces a detailed analysis of papers published in some economics journals. Section \ref{sec:futureReseach} identifies literature gaps and explores open research lines. Finally, Section \ref{sec:Conclusions} draws the main conclusions.

\section{Methodological design}
\subsection{First step: distant reading by means of bibliometric analysis\label{sec:bibliometric}}
Our first approach to this survey is to extract articles' metadata from Web of Science Core Collection (WoS), Clarivate Analytics. We conducted the following query: 

\texttt{ALL=(bitcoin OR ethereum OR litecoin OR monero OR iota) NOT AU=(Iota) AND WC=(Business OR Business, Finance OR Economics)}

We retrieved papers from all the years included in the core collection of the Web of Science, which gave a total of 626 papers. We restrict our sample only to articles, which means that we discard conferences proceedings and book chapters. This amounts 444 articles. Finally, we take out of our sample articles published in Forbes. The reason is that Forbes has a great impact among practitioners, CEOs, and general public, but it is seldom cited in scientific publications. Thus, the total number of articles in our bibliometric analysis is 438. The analysis of this section was conducted using bibliometrix R package, developed by \cite{bibliometrix}. The detail of the top sources is displayed in Table \ref{tab:sources}.

\begin{table}
  \centering
  \caption{Most frequent sources}
    \begin{tabular}{rlr}
    \toprule
    \multicolumn{1}{l}{\#} & Sources & \multicolumn{1}{l}{Articles} \\
    \midrule
    1     & Finance Research Letters & 56 \\
    2     &  Economics Letters & 42 \\
    3     & Journal Of Risk and Financial Management & 21 \\
    4     & Research in International Business and Finance & 20 \\
    5     & International Review Of Financial Analysis & 16 \\
    6     & Applied Economics Letters& 15 \\
    7     & Applied Economics  & 12 \\
    8    & Journal Of Risk Finance & 9 \\
    9     & Economics Bulletin & 6 \\
		10    & Journal of International Financial Markets Institutions \& Money & 6\\
    11     & Quarterly Review Of Economics And Finance & 6 \\
    12    & Journal of Corporate Accounting and Finance & 5 \\
		13 &   North American Journal of Economics and Finance & 5 \\
    14    & Annals Of Financial Economics & 4 \\
		15   & Business Horizons  & 4 \\
    16    & Financial Innovation & 4 \\
    \bottomrule
    \end{tabular}%
  \label{tab:sources}%
\end{table}%

Our sample contains 38 Highly Cited Papers (HCP)\footnote{Highly Cited Papers is a metric developed by Web of Science Group, to help to identify top-performing research. 
HCP are papers that have received enough citations to place them in the top 1\% when compared to all other papers published in the same year in the same field. For additional details of this and other metrics see: \url{https://clarivate.libguides.com/esi}.}. Among all HCP, 15 were published in \textit{Economics Letters}, and 12 in \textit{Finance Research Letters}.

Our bibliometric analysis identified the most cited papers. We detect that 4 and 6 out of the 20 most cited were  published in Finance Research Letters and Economics Letters, respectively (see Table \ref{tab:citations}).

\begin{table}
  \centering
  \caption{Top 20 manuscript per citations}
    \begin{tabular}{lrr}
    \toprule
    Paper & \multicolumn{1}{l}{Total Citation} & \multicolumn{1}{l}{Citation per year} \\
    \midrule
    \cite{Boehme2015} & 198   & 39.6 \\
    \cite{URQUHART201680} & 179   & 44.8 \\
    \cite{CHEAH201532} & 164   & 32.8 \\
    \cite{DYHRBERG201685} & 154   & 38.5 \\
    \cite{KATSIAMPA20173} & 120   & 40 \\
    \cite{Dwyer2015} & 119   & 23.8 \\
    \cite{BOURI2017192} & 118   & 39.3 \\
    \cite{Ciaian2016}& 116   & 29 \\
    \cite{NADARAJAH20176} & 110   & 36.7 \\
    \cite{DYHRBERG2016139} & 108   & 27 \\
    \cite{Bariviera20171} & 95    & 31.7 \\
    \cite{CORBET201828} & 85    & 42.5 \\
    \cite{BALCILAR201774}& 84    & 28 \\
    \cite{BaekElbeck2015} & 77    & 15.4 \\
    \cite{URQUHART2017145} & 68    & 22.7 \\
    \cite{BaurHongLee2018} &   66 & 33 \\
    \cite{BOURI201787} & 65    & 21.7 \\
    \cite{CheungRoca2015} & 64    & 12.8 \\
    \cite{Selgin2015} & 59    & 11.8 \\
    \cite{FRY2016343} & 58    & 14.5 \\
    \bottomrule
    \end{tabular}%
  \label{tab:citations}%
\end{table}%

Finally, the analysis of authors keywords and Keyword-Plus\footnote{Keyword-Plus are those extracted from the titles of the cited references by Thomson Reuters (the company maintaining WoS). Keyword Plus are automatically generated by a computer algorithm.}, allows to detect the main topics of papers in our sample. This keywords helped to form the groups developed in the following section. Both groups of keywords, indicate that: (i) bitcoin seems to be the predominant object of the studies, (ii) most words are finance-related, (iii) there are clusters of literature devoted to informational efficiency, safe haven condition, volatility, hedge properties, and price bubbles.

\begin{table}
  \centering
  \caption{Most relevant keywords.}
    \begin{tabular}{lrlr}
    \toprule
    Author Keywords (DE) & \multicolumn{1}{l}{Articles} & Keywords-Plus (ID) & \multicolumn{1}{l}{Articles} \\
    \midrule
    Bitcoin & 257   & Bitcoin & 101 \\
    Cryptocurrency & 124   & Inefficiency & 79 \\
    Cryptocurrencies & 75    & Volatility & 65 \\
    Blockchain & 47    & Economics & 49 \\
    Volatility & 23    & Gold  & 49 \\
    GARCH & 17    & Hedge & 40 \\
    Digital Currency & 15    & Returns & 34 \\
    Ethereum & 15    & Safe Haven & 23 \\
    Market Efficiency & 15    & Dollar & 20 \\
    Safe Haven & 13    & Exchange & 20 \\
    Money & 10    & Market & 19 \\
    Crypto Currency & 9     & Time Series & 18 \\
    Gold  & 8     & Prices & 17 \\
    Hedge & 8     & Currency & 15 \\
    Virtual Currency & 8     & Money & 15 \\
    Forecasting & 7     & Cryptocurrencies & 14 \\
    Long Memory & 7     & Markets & 14 \\
    Bubbles & 6     & Unit Root & 14 \\
    Commodities & 6     & Model & 13 \\
    Distributed Ledger & 6     & Models & 13 \\
    \bottomrule
    \end{tabular}%
  \label{tab:keywords}%
\end{table}%

\subsection{Second step: close reading of cryptocurrency literature\label{sec:literature}}

Bibliometric analysis conducted in the previous section, shows main characteristics of the data set. However it has two drawbacks. First, although powerful machine learning techniques are used, bibliometric analysis is not a substitute, but rather a complement of a comprehensive literature review. Second, papers included in Web of Science experience a time delay to be introduced into the database. There are numerous accepted papers that published online in their respective journal websites, but they are not yet indexed in Web of Science.

Considering this situation, based on the previous bibliometric analysis we conduct a close reading of all the papers (including articles in press), from the two most frequent journals (Economics Letters and Finance Research Letters), and the International Review of Financial Analysis. The reason for this selection is twofold. On the one hand, 26\% of the papers on cryptocurrencies have published in these journals. On the other hand, 30 out of the 38 Highly Cited Papers in this area are published in these three journals. Then, we can say that mainstream research of this topic is conveyed around these three journals. Additionally, we include in our analysis the papers by \cite{Boehme2015} and \cite{Gandal2018}, published in the {\it Journal of Economic Perspectives} and in the {\it Journal of Monetary Economics}, respectively, because they are the only papers published in journals classified at level 4 (world-wide exemplars of excellence) by the \cite{CABS}.

\section{Close reading findings}

The dataset in this section is different from the one used in Section \ref{sec:bibliometric}. Out of the 116 articles  published in ,{\it Economics Letters},{\it Finance Research Letters}, {\it International Review of Financial Analysis}, {\it Journal of Monetary Economics}, and {\it Journal of Economic Perspectives}, we selected 98 articles. The distribution of papers read per source is detailed in Table \ref{tab:articlesRead}.
A meticulous analysis of each paper, detailing cryptocurrencies studied, data frequency, source of data, quantitative methodology, aim of the paper and main results, is displayed in Table \ref{tab:detailPapers} in the Appendix. In the following subsections we will highlight the salient features of some representative papers.

\cite{Bohme2015} is one of the earliest papers to render a full overview of bitcoin and its relationship with the then emerging blockchain technology. The authors point out pros and cons of bitcoin, emerging challenges for the monetary policy, risks, and necessity of regulation. It constitutes an excellent introductory paper, in order to begin the study of this field. 

\begin{table}
\centering
\caption{Publication sources considered in our sample}
\begin{tabular}{lrr}
\toprule
Journal	&\# articles	&\% \\
Economics Letters	&33&	34\% \\
Finance Research Letters	&49&	50\% \\
International Review of Financial Analysis&	14	&14\% \\
Journal of Monetary Economics & 1& 1\% \\
Journal of Economic Perspectives & 1 & 1\% \\
Total	& 98& 100\%\\	
\bottomrule
\end{tabular}
\label{tab:articlesRead}
\end{table}

\subsection{Data sources}
Our first analysis is related to the source of data used in papers. Table \ref{tab:datasources} displays the data sources used in the papers of our sample. We detect that 61\% of the papers use data from either Coinmarketcap, Coindesk, or Bitcoincharts. One of the reasons is, apparently, that these websites allow the use of Application Programming Interfaces (API). An API is a set of subroutine definitions and communication protocols that allow, among other things, to formulate data requests, and download data in an efficient way. In addition, all three websites gather information from several trading platforms and several cryptocurrencies. Thus, they provide a broad coverage of the market. With the exception of three papers, the rest rely on only one source of data. 

Considering that these websites generate their own price indices by averaging different cryptocurrencies' platforms, data are not homogeneous across all papers. This situation emerges as a weakness in order to compare results. It is well known in financial economics, that equally-weighted indices or capitalization-weighted indices can lead to different results in stock markets. A similar situation can happen in the cryptocurrency market. Special attention should be payed to the use of nontraded prices or non-synchronous data in multivariate analysis.  A very recent and detailed critical review of cryptocurrency data is in \cite{Alexander2020}, where it is reported that half of the papers published since 2017 uses appropriate data.

\begin{table}
\centering
\caption{Source of data used in empirical studies of cryptocurrencies}
\begin{tabular}{rrr}
\toprule
Source &\# articles &\%\\
\midrule
Coinmarketcap & 27 &26\%\\
Coindesk & 20 & 19\%\\
Bitcoincharts &13 &13\%\\
Other & 36 & 35\%\\
Not known & 5 &5\%\\
Not applicable & 2 & 2\% \\
Total &103$^*$ &100\%\\
\bottomrule
\multicolumn{3}{l}{$^*$ Total of articles does not match}\\ 
\multicolumn{3}{l}{because some papers use more than one source}\\
\label{tab:datasources}
\end{tabular}
\end{table}

\subsection{Data frequency}

An important issue in our literature review, is to detect the data frequency used in the empirical studies. Unlike stock or bond markets, cryptocurrencies markets offer free, real time information. Moreover, trading is open 24/7. From a theoretical point of view, if the goal is to understand a stochastic process, recorded in a time series, sampling selection is a key task. In this sense, cryptocurrencies (specially the bigger ones) offer the possibility to select different data granularity. We detect that the large majority of empirical studies (81\%) uses daily data, whereas intradaily data is only used by 14\% of the papers. It seems that authors consider daily frequency as the ``natural frequency'' of data, disregarding other options. This situation means that there are still unexplored issues, which could give new insights and possible uncover stylized facts at ultra-high frequency. 

\begin{table}
\centering
\caption{Data frequency used in empirical studies of cryptocurrencies}
\begin{tabular}{lrr}
\toprule
Data frequency	& \# articles &\% \\
\midrule
Daily	 & 79 & 81\%\\
Intraday &	13 & 14\%\\
Weekly & 	3 & 3\%\\
Monthly  &1 & 1\%\\
Not known/not applicable&2&1\%\\
Total &98  & 100\%\\
\bottomrule
\end{tabular}
\label{tab:frequency}
\end{table}

\subsection{Main research topics}
After a detailed reading of the 98 papers in our sample, we classify them according to their key research topics.  Even though some papers cover more than one topic, we assign the one that, in our opinion, is the main driver of their research. In the following subsections, we select some articles of each research topic in order to explain the methodologies and main findings. 

Classification and detailed characteristics of all 98 papers are displayed in Table \ref{tab:detailPapers} (Appendix). Almost half of them are referred either to classical financial economics topics such as informational efficiency (26\%), price discovery (16\%), or price clustering (3\%). There is another portion of literature that studies the characteristics of volatility (15\%). Another important research line goes along two related topics: portfolio formation (11\%) and safe-haven properties of cryptocurencies (7\%). There is only one paper that performs a literature review in our sample\citep{Corbet2019}, whose coverage only partially overlaps with ours.

\begin{table}
\centering
\caption{Articles' key research topics}
\begin{tabular}{rrr}
\toprule
Research topic	& \# articles  &	\%\\
Informational efficiency	&25&	26\%\\
Price discovery	&15&	16\%\\
Volatility&	13	&13\%\\
Portfolio formation	&10	&10\%\\
Bubble	&8&	8\%\\
Safe-haven&	7&	7\%\\
Correlation &7&	7\%\\
Microstructure	&6&	6\%\\
Price clustering	&3&	3\%\\
Monetary economics	 &2&	2\%\\
Literature review	&1&	1\%\\
Overview  & 1  & \% \\
Total &	98&	100\%\\
\bottomrule
\end{tabular}
\end{table}

\subsubsection{Monetary economics and overview of bitcoin ecosystem}
Papers in this section conducts general analysis of bitcoin prices and demand, giving an overview of the functioning of this new kind of financial market. 
\cite{Gandal2018} identifies and analyzes the impact of suspicious trading activity on one important trading platflorm, concluding that cryptocurrency markets are vulnerable to manipulation due to the unregulated nature of the activity. Recently, \cite{DELAHORRA201921} focus their analysis on the determinants of the demand for bitcoin, building monetary-theory based demand model. They find that, in the short run, speculation fuels the demand for bitcoin. However, in the long run demand is driven by expectations about its future utility as a medium of exchange.

\subsubsection{Informational Efficiency \label{sub:EMH}}
There is a relevant number of papers inquiring on the informational efficiency of cryptocurrencies.  Articles within this group are aimed at testign the weak form of the Efficient Market Hypothesis (EMH), developed by \cite{Fama70}, which states that prices in an informational efficient market should follow a random walk. The three most cited within this group are published in the {\it Economics Letters}. Although some of the articles from other groups also study some characteristics dealing with the efficiency of cryptocurrencies, some difference between them are found.

The methodology used by \cite{URQUHART201680}, the highest cited article in this group, to test the EMH has been used subsequently in other articles. In that article, a battery of tests for randomness are employed:
\begin{itemize} 
\item \cite{LjungBox1978} test, in order to test the null hypothesis of no autocorrelation.
\item \cite{WaldWolfowitz1940} and \cite{Bartels1982} tests to determine whether returns are independent.
\item Variance ratio test by \cite{LoMacKinlay1988}, which under the null hypothesis, the price process is a random walk. Papers also use some variations such as the automatic variance test (AVR) by \cite{Choi1999}, or the the wild-bootstrapped version by Kim (2009).
\item \cite{BDS1996} test, in order to verify possible deviations from independence including linear dependence, non-linear dependence, or chaos.
\item \cite{Hurst1951} Rescaled Hurst exponent (R/S Hurst) to detect the presence of long memory in prices time series.
\end{itemize}
\cite{URQUHART201680} finds that Bitcoin had been informational inefficient at the beginning, but was moving towards a more efficient market.

\cite{NADARAJAH20176} use, in addition to the previous tests, the following ones:
\begin{itemize} 
\item Spectral shape tests by \cite{Durlauf1991} and \cite{Choi1999} to test for random walk. 
\item	\citep{EscancianoLobato2009} robustified portmanteau test for no serial correlation.
\item Generalized spectral test by \citep{EscancianoVelasco2006} to check whether the martingale difference hypothesis holds for the returns.
\end{itemize}
In this paper, the authors show that some power transformations of Bitcoin returns can be weakly eficient.

Additionally, \cite{Bariviera20171} compares results of the Hurst exponent computed by R/S and Detrended Fluctuation Analysis (DFA) methods. The author argues in favor of the latter because it avoids the spurious detection of long-range dependence. The main contribution of this paper is to study daily returns and volatility using sliding windows. Such methodology design allows detecting a diminishing memory in daily returns, but persistent memory in volatility, justifying the use of GARCH modelization in variance. 

\subsubsection{Price discovery}
The articles from this group employ different approaches to study the predictability of cryptocurrencies. For example, some papers apply machine learning algorithms in order to measure the forecasting power of past Google or Twitter searches. 

\cite{BRAUNEIS201858} uses the EMH tests introduced by \cite{URQUHART201680} as measure of how predicable cryptocurrencies are. Furthermore, they also add a Measure Of Efficiency (MOE) \citep{Godfrey2017}, using different kind of liquidity measures. MOE measures how well a passive strategy performs relative to active trading. The four liquidity measures proposed are the following: (1) log-dollar volume, (2) turnover ratio, (3) Amihud's illiquidity ratio \citep{AMIHUD200231} and (4) bid-ask estimate \citep{Corwin2012}. 

Moreover, \cite{URQUHART201840} constructs a time series of daily realized volatility (RV), which was introduced by \cite{Andersen2003}. This model is built using vector autoregressive model (VAR) to study the dynamics between search queries (Google Trends data), realized volatility, trading volume and returns. \cite{URQUHART201840} finds that attention of Bitcoin is significantly influenced by the previous day's high realized volatility and volume.

In addition, \cite{AALBORG2019255} use four OLS models to study returns, volatility and trading volume of Bitcoin. Some of the independent variables are the trading volume, VIX index, Google trends data, etc. To study the volatility, they use the HAR-RV model proposed by \cite{Corsi2009}, to capture long-memory behavior of volatility.
The authors present alternative models using: (1) daily data, (2) daily data and lagged independent variables, (3) weekly data, (3) weekly data and lagged independent variables.
\cite{AALBORG2019255} find that none of the considered variables can predict Bitcoin returns and the trading volume of Bitcoin can be predicted from Google searches for Bitcoin. 

\subsubsection{Price volatility \label{sec:volatility}}
Cryptocurrencies are highly volatile (approximately ten times more than traditional assets), due to the intrinsic speculative characteristics of the investments, the velocity of transactions, and the unregulated environment. The group of articles under this label study some stylized facts of the volatility of returns of the cryptocurrencies. Most of the articles of this group, based on previous experience in other financial markets, apply different variations of GARCH models. This type of models are suitable for estimating the time-varying volatility. Most papers find volatility clustering, which implies that there are periods of relative calm followed by periods of swings. This fact is also known as persistence of the volatility.

\cite{KATSIAMPA20173} compares different first order GARCH-type model for the conditional variance, with an autoregressive model for the conditional mean. Particularly, the applied models are: GARCH, EGARCH, TGARCH, APARCH, CGARCH and ACGARCH. It is found that the optimal model is the AR-CGARCH model, which suggests the importance of having both a short-run and a long-run component of conditional variance.

\cite{ARDIA2019266} is an extension of \cite{KATSIAMPA20173}. The model used is a Markov–switching GARCH (MSGARCH) to capture any regime changes in the Bitcoin volatility dynamics, and outperform single-regime GARCH specifications in Value-at-Risk (VaR) forecasting. 

\cite{Katsiampa2018} studies the volatility dynamics of the two major cryptocurrencies (Bitcoin and Ether), using a bivariate GARCH (BEKK model). Her results suggest that price returns of both cryptocurrencies are stationary, but exhibit volatility clustering. 

Finally, \cite{GKILLAS2018109} use extreme value theory to investigate tail behavior in cryptocurrencies. In particular, they study the two major tail risk measures of VaR and Expected Shortfall (ES) as extreme quantiles of the Generalized Pareto distribution (GPD). They apply a parametric bootstrap bias-correction approach to the two risk measures in order to reduce any uncertainty resulting from the estimation procedure of the asymptotic extreme value distribution and the threshold selection. This study tells the different degree of riskiness of each cryptocurrency under examination.

\subsubsection{Assets correlation and portfolio optimization \label{sec:CorrPortolio}}
This group of articles study the relationship between cryptocurrencies and the other assets. The objective of these articles is to compare the behavior of cryptocurrencies with respect to traditional assets and to evaluate the possibility of adding cryptocurrencies to current financial portfolios. In addition, some papers explore the suitability of constructing cryptocurrency-only portfolios. The rationale is that, due to the low correlation of cryptocurrencies \textit{vis-\`a-vis} traditional assets, they can reduce the risk of the overall portfolio. Most of the studies suggest that cryptocurrencies can become a portfolio diversifier. However, most authors warn that it is important to evaluate the uncertainties around future regulation and the exposure of cryptocurrencies to hacking activities.

\cite{DYHRBERG201685} applies GARCH models to determine that bitcoin has a place on the financial markets and in portfolio management, as it can be classified as something in between gold and the US dollar. Nevertheless, \cite{BAUR2018103} replicated this study proving that Bitcoin exhibits distinctively different return, volatility and correlation characteristics compared to other assets, including gold and the US dollar. \cite{BAUR2018103} extends \cite{DYHRBERG201685}, adding the asymmetric GARCH model to the analysis.

In addition, \cite{GUESMI2019431} implement various specifications of the DCC-GARCH models to investigate volatility spillovers between Bitcoin and exchange rates, stock market, and commodity series.  They find that VARMA (1,1)-DCC-GJR-GARCH is the best model specification to describe the joint dynamics of Bitcoin and different financial assets. This suggests that Bitcoin may offer diversification and hedging benefits for investors.

In another vein, \cite{LIU2019200} considers different portfolio models (1/N equal weighted (EW), minimum variance (MV), risk parity (RP), Markowitz (MW), maximum Sharpe ratio (MS), and maximum utility (MU)) to examine the investability and diversification benefits of cryptocurrencies. This author shows that portfolio diversification across different cryptocurrencies can significantly improve the investment results.

Lastly, \cite{CORBET201828} examine the relationships between three popular cryptocurrencies (Bitcoin, Litecoin, Ripple) and a variety of traditional financial assets. They use the generalized variance decomposition methodology by \cite{DIEBOLD201257} to estimate the direction and intensity of spillovers across selected markets. Furthermore, they estimate unconditional connectedness relations in time-frequency domain (\cite{Barunik2016}). They find evidence of the relative isolation of these assets from the financial and economic assets. \cite{Aslanidis2019}, using a generalized DCC model \citep{Engle2002339}, find similar results to \cite{CORBET201828}, and also uncovers that crosscorrelation against Monero is more stable across time that other correlation pairs.

\subsubsection{Safe-haven characteristics}
Related to the previous category, articles dealing with safe-have characteristics evaluate if bitcoin can become a substitute for gold. The rationale behind this group of articles is that both are uncorrelated with other financial assets. 

Some papers in this section upholds that cryptocurrencies are not only useful portfolio diversifiers but also ``wealth shields''. Therefore, authors consider cryptocurrencies a commodity, rather than a medium of exchange. However, as explained in section \ref{sec:CorrPortolio}, the doubts around their regulations, the lack of security due to cyberattacks, the enormous volatility (see section \ref{sec:volatility}) and the lower liquidity (compared to traditional assets) still generate uncertainty around cryptocurrencies as safe-haven assets.

\cite{DYHRBERG2016139} finds some relationship between bitcoin and gold. This paper uses the threshold GARCH (TGARCH) model \citep{glosten1993} to examine if bitcoin could be used as a hedge against stocks in the Financial Times Stock Exchange index (FTSE) and the US dollar. The author affirms taht bitcoin possess some of the same hedging abilities as gold. In the same vein, \cite{BOURI201787} investigate whether bitcoin can hedge global uncertainty, measured by the first principal component of the VIXs of 14 developed and developing equity markets. They use the wavelet transform to decompose bitcoin returns into its various frequencies (or investment horizons). Their results show that hedging for bitcoin is observed at shorter investment horizons, and at both lower and upper ends of bitcoin returns and global uncertainty.

Conversely, some of the recent papers disagree with this view of bitcoin becoming a hedge or a safe-haven asset. For example, \cite{Klein2018} use different GARCH models (including BEKK-GARCH) to show that bitcoin does not reflect any distinctive properties of gold other than asymmetric response in variance. Moreover, they show that FIAPARCH is be the best fitting model in terms of log-likelihood and information criteria. Furthermore, \cite{Smales2018} argues that it is unlikely to be worthwhile considering  bitcoin as a safe haven asset because is more volatile, less liquid, and costlier to transact (in terms of time and fees) than other assets (including gold), even in normal market conditions. \cite{BOURI2017192} show, using the Bivariate Dynamic Conditional Correlation (DCC) model by \cite{Engle2002339}, that bitcoin can usually serve as an effective diversifier but it has only hedge and safe haven properties against Asia Pacific stocks.

\subsubsection{Bubble formation}

Bubble behavior of cryptocurrencies easily captures media attention. This fact is one of the main drivers that made cryptocurrencies (mainly bitcoin) famous for most of the people in 2017 . Therefore, in this group of articles different empirical tools are used to study the bubble behavior of cryptocurrency prices

\cite{CHEAH201532} empirically estimate bitcoin's fundamental value. They use the Intrinsic Rate of Return and the Intrinsic Level of Risk measures. Moreover, they use the bubble models by \cite{johansen2000}, \cite{ANDERSEN2004565}, and \cite{MacDonell2014}. They show that bitcoin exhibits speculative bubbles even before the big bubble of 2017. Furthermore, they find empirical evidence that the fundamental price of bitcoin is zero, which raises serious concerns upon the long-term sustainability of bitcoin.

Later, the same authors (\cite{FRY2016343}) developed probabilistic and statistical formulation of econophysics models to test for economic bubbles and crashes. They use three estimations. Firstly, the univariate and negative bubbles \citep{johansen2000,YAN20121361}. Secondly, multivariate models that describe the price of more-than-one asset simultaneously and are significant for empirical applications. Thirdly, a bivariate bubble model, which is a method to test for the presence or absence of contagion during bubbles and negative bubbles. In addition, they also examine unpredictable market shocks. They find evidence of a negative bubble from 2014 onwards in the two largest cryptocurrency markets, bitcoin and ripple. Furthermore, evidence suggests that there is a spillover from ripple to bitcoin that exacerbate price falls in bitcoin.

Finally, \cite{BOURI2019178} test the co-explosivity of cryptocurrencies. This paper is the first to study co-explosivity (that is, the potential interactions among bubble periods within the cryptocurrency market). The methodology used is the generalized supremum Augmented Dickey–Fuller (GSADF) test of \cite{Phillips2015} and a logistic regression to uncover evidence of co-explosivity across cryptocurrencies. They find evidence of a multidirectional co-explosivity behavior that is not necessarily from bigger to smaller and younger markets.

\section{Literature gaps and open research paths \label{sec:futureReseach}}
According to our review, most of the papers regarding cryptocurrencies are focused on financial aspects of cryptocurrencies: informational efficiency, volatility, portfolio optimization, bubble behavior, etc. 

The cryptocurrency market, unlike traditional assets, are opened 24/7. We can find trades taken place almost every minute for the most liquid cryptocurrencies. Then, this market offers a unique opportunity to test continuous time models, that can be hardly verified in traditional stock or bond markets. 

As shown in Table \ref{tab:frequency}, most papers are focused on daily data. Probably this is a customary use from financial economists when studying stock markets. However, it would important to explore the information gain (if it exists) in the use of high frequency data. In addition, considering cryptocurrencies as pure speculative assets, their study at high frequency could give some hints on the behavior of traditional assets whose behavior at high frequency cannot be observed.

One topic, usually developed in engineering journals is the environmental impact of cryptocurrencies' mining. This theme is mostly not yet studied in economics journals. Even when authors may comment on the important electricity consumption of cryptocurrencies during the mining process, they fail to make a clear estimation of the environmental impact of blockchain technology as a whole. In other words, there is a need for an analysis of positive and negative externalities of the blockchain technology.

Another gap in the literature is how mining protocols could affect price. It is well known that cryptocurrencies use different protocols to maintain network consensus\footnote{For example, bitcoin uses `proof of work', DASH and NEO use `proof of stake', Burstcoin uses `proof of capacity', etc. There are other alternative protocols, e.g.  proof of authority, proof of space. For a recent discussion of these and other technical aspects see \cite{Belotti20193796}}. To the best of our knowledge there is no paper considering the influence of consensus protocols in price formation, returns or volatility.

Additionally, we detect that there is a lack of theoretical papers that contemplate the potential impact of national (or even supranational) regulation in this market. It is remarkable the lack of an institutional economics view of this phenomena.

Finally, as we highlight in this paper, most past research was focused exclusively on bitcoin, or at most in the four or five most important cryptocurrencies. Even though bitcoin represents approximately 68\% of the market capitalization in January 2020, there are currently more than five thousands active cryptocurencies \citep{coinmarketcap}. Extending previously used models to more cryptocurrencies can give more information about this market as a whole, putting together assets with different underlying technology, liquidity, different age, etc.

\section{Conclusions \label{sec:Conclusions}}

This study makes a bibliometric and literature review of the most important economic topics studied on cryptocurrencies. Bibliometric studies are a useful technique to analyze the state of the art in an specific field with large number of papers, because it could be processed by means of machine learning algorithms. However, it could hardly substitute the insight given by the specialized researcher. Consequently, our methodology is based on a combination of machine learning (for bibliometric analysis), and close reading (for literature review). The first step allows for an informed sample selection of papers, which is used in the second step. This literature review has a dual goal. First, to propose this hybrid methodology. Second, to provide an updated, useful review for new and experienced researchers in this field. 

Our analysis displayed the main research lines, and some emerging paths of this novel market. We expanded previous literature, adding a comprehensive review of 98 papers, classifying them into different research topics, and identifying top papers and journals. Finally we detected some literature gaps and propose future research paths.

\bibliographystyle{apalike}
\bibliography{bibliometrics}

\section{Appendix}

\newgeometry{top=.5in,bottom=.5in,right=0.5in,left=0.5in}
\scriptsize

\begin{landscape}

\begin{longtable}{ p{30mm} p{20mm} p{20mm} p{20mm} p{18mm}  p{40mm} p{40mm} p{40mm}}

\caption{Detailed analysis of papers selected in Section \ref{sec:literature}}\\
\centering 


Paper & Group & Cryptocurrencies studied & Data Frequency & Source of data & Methodology & Aim of the paper & Results \\ \midrule
\endhead
\hline \multicolumn{4}{r}{\textit{Continued on next page}} \\
\endfoot
\cite{CHEAH201532} & Bubble & Bitcoin & Daily & Coindesk & MacDonell (2014) test for bubbles, model in Johansen et al. (2000), model in Andersen and Sornette (2004) & Provide empirical evidence to address the existence of bubbles in Bitcoin markets. Determine the fundamental value of Bitcoin & Bitcoin exhibits speculative bubbles. The fundamental price of Bitcoin is zero \\ 
\cite{FRY2018225} & Bubble & Bitcoin, Ripple, Ethereum, Bitcoin Cash & Daily & Coin\-mar\-ket\-cap & Theoretical refinement of the model in Cheah and Fry (2015)* & Develop rational bubble models & Evidence of bubbles in Bitcoin and Ethereum. No evidence of bubbles in Ripple once we account for heavy tails and liquidity risk. \\ 
\cite{BOURI2019178} & Bubble & Bitcoin, Ripple, Ethereum, Litecoin, Nem, Dash, Stellar & Daily & Coin\-mar\-ket\-cap & GSADF, Logistic regression & Data-stamp price explosivity in leading cryptocurrencies & Cryptocurrencies  characterised by multiple explosivity. Multidirectional co-explosivity behaviour that is not necessarily from bigger to smaller and younger markets \\ 
\cite{GEUDER2018} & Bubble & Bitcoin & Daily & Coin\-mar\-ket\-cap & PSY (SADF, GSADF), LPPL & Study bubble behavior in Bitcoin prices during 2016-2018 & Bubble behavior is a common and reoccurring characteristic \\ 
\cite{CORBET201881} & Bubble & Bitcoin, Ethereum & Daily & API & Phillips et al. (2011) (SADF, GSADF) & Examine the existence and dates of pricing bubbles in Bitcoin and Ethereum & There are periods of clear bubble behavior, with Bitcoin in Nov. 2017 almost certainly in a bubble phase \\ 
\cite{CAGLI2019398} & Bubble & Bitcoin, Ethereum, Ripple, Litecoin, Stellar, Nem, Dash, and Monero & Daily & Coin\-mar\-ket\-cap & Multi-equation continuous time system & Investigate explosive behavior & Almost all cryptocurrencies exhibit explosive behavior and significant pairwise co-movement \\ 
\cite{FRY2016343} & Bubble & Bitcoin, Ripple & Daily & Co\-in\-desk, Coin\-mar\-ket\-cap & Univariate and bivariate bubbles, multivariate models & Develop a suite of models for financial bubbles and crashes & Negative bubble from 2014 onwards in Bitcoin and Ripple\\
\cite{GKILLAS2018109} & Bubble &Bitcoin, Ethereum, Ripple, Bitcoin Cash, Litecoin
  & Daily & Coindesk, Coin\-mar\-ket\-cap & Extreme value analysis & Study the tail behavior of the returns
 & Bitcoin Cash is the riskiest cryptocurrency, while Bitcoin and Litecoin are the least risky.\\
\cite{CORBET201828} & Correlation & Bitcoin, Ripple, Litecoin & Daily & Cryp\-to\-com\-pa\-re & GVD, BK & Analysis of  crosscorrelation of crypto and traditional assets over short and long horizons & Relative isolation of cryptos from traditional assets \\ 
\cite{Aslanidis2019} & Correlation & Bitcoin, Ripple, Dash, Monero & Daily & Coin\-mar\-ket\-cap & generalized DCC & Analysis of  crosscorrelation of crypto and traditional assets  & cryptocurrencies exhibit similar mean correlation among them, and detached from traditional assets. Monero correlations are more stable\\ 
\cite{KOUTMOS2018122} & Correlation & 18 cryptocurrencies & Daily & Coin\-mar\-ket\-cap & VAR, spillover index & Measure return and volatility spillovers among cryptocurrencies & Growing interdependence among cryptocurrencies, being Bitcoin the dominant transmitter of shocks\\ 
\cite{WEI201819} & Correlation & Bitcoin, Tether & Daily & Coin\-mar\-ket\-cap & ADL Granger causality, VAR & Examine the impact of cryptocurrency issuances on cryptocurrency returns & Tether grants were potentially timed to follow Bitcoin downturns and subsequent Bitcoin/Tether trading volumes increased \\ 
\cite{TU2018} & Correlation & Bitcoin, Litecoin & Daily & Coin\-mar\-ket\-cap & Granger causality, BEKK-MGARCH & Study the effect of the bifurcation of Bitcoin on its interactions with Litecoin & Bifurcation weakened the market position and pricing influence of Bitcoin \\ 
\cite{WANG2018} & Correlation & Bitcoin & Daily & Coindesk & MVQM, Granger causality & Investigate risk spillover effect from economic policy uncertainty (EPU) to Bitcoin & Risk spillover effect from EPU to Bitcoin is negligible \\ 
\cite{GIUDICI2019309} & Correlation & Bitcoin & Daily & Some exchanges & Network VAR & Understand price transmition between different crypto market exchanges, and between crypto and traditional assets & Correlation  between bitcoin prices exchanges is strong, correlation of bitcoin prices with traditional assets is low \\ 
\cite{URQUHART201680} & Efficiency & Bitcoin & Daily & Bit\-coin\-a\-ve\-ra\-ge & LB, runs test, Bartels, VR, AVR, WBAVR, BDS, Hurst exponent & Study the informational efficiency of Bitcoin & Bitcoin in an inefficient market but  moving towards an efficient market \\ 
\cite{NADARAJAH20176} & Efficiency & Bitcoin & Daily & Bit\-coin\-a\-ve\-ra\-ge & LB, runs test, Bartels, WBAVR, SST, BDS, RPT, GS & investigate the market efficiency of Bitcoin & A power transformation of Bitcoin returns can be weakly efficient \\ 
\cite{Bariviera20171} & Efficiency & Bitcoin & Daily & Datastream & Hurst exponent (R/S, DFA) & Study long-range dependence of Bitcoin return and volatility & Daily return time series become more efficient across time. Daily volatility exhibits long-range memory\\ 
\cite{PHILLIP20186} & Efficiency & 224 cryptocurrencies & Daily & Brave New Coin (BNC)  & GLM, SV, Leverage, Heavy tails & Measure and compare the varied nature of cryptocurrencies & Cryptocurrencies exhibit long memory, leverage, stochastic volatility and heavy tailedness. \\ 
\cite{KHUNTIA201826} & Efficiency & Bitcoin & Daily & Coindesk & DL, GS, AMH & Evaluate the adaptive market hypothesis (AMH) in Bitcoin market & The evidence of dynamic efficiency  \\ 
\cite{VANVLIET201870} & Efficiency & Bitcoin & Monthly & Block\-cha\-in.\-in\-fo & Metcalfe's Law & Present new model of the market capitalization of Bitcoin built upon Metcalfe's Law & Model fits empirical data well \\ 
\cite{TIWARI2018106} & Efficiency & Bitcoin & Daily & Coindesk & DFA, CMA-1, CMA-2, Periodogram-LAD, Periodogram-LS, GPH, and MLE techniques & Revisit the issue of informational efficiency of Bitcoin & The market is informational efficient \\ 
\cite{WEI201821} & Efficiency & 456 cryptocurrencies & Daily & Coin\-mar\-ket\-cap & LB, Bartels, VR, AVR, BDS, Hurst exponent, AIR & Examine the liquidity of 456 cryptocurrencies & Return predictability diminishes as liquidity increases in cryptocurrencies \\ 
\cite{CHEAH201818} & Efficiency & Bitcoin & Daily & Bit\-coin\-charts & FCVAR, Log periodogram, ELW & Test whether cross-market Bitcoin markets display heterogeneous informational inefficiency & Evidence of long-memory in individual markets and the system of markets \\ 
\cite{TAKAISHI20185} & Efficiency & Bitcoin & Intraday (1-minute) & Coindesk & Autocorrelation function & Investigate the Taylor effect in Bitcoin time series & The Taylor effect is present in Bitcoin time series \\ 
\cite{KOCHLING201939} & Efficiency & 75 cryptocurrencies & Daily & Coin\-mar\-ket\-cap & Delay measures proposed by \cite{HouMoskowitz} & Investigate the reaction time to unexpected relevant information & Average price delay significantly decreases during the last three years. Price delay is highly correlated to market capitalization and liquidity \\ 
\cite{THIES2018223} & Efficiency & Bitcoin & Daily & Bitstamp & Bayesian change point model & Study existence of structural breaks in the average return and volatility of the Bitcoin price & Structural breaks in average returns and volatility of Bitcoin are very frequent \\ 
\cite{AHARON2018} & Efficiency & Bitcoin & Daily & Bit\-coin\-charts & OLS, GARCH, QMLE & Extend the exploration of the day-of-the-week effect to Bitcoin & Evidence about  day-of-the-week effect anomaly in returns and volatility\\ 
\cite{CHEVAPATRAKUL2018} & Efficiency & Bitcoin & Daily & Coin\-mar\-ket\-cap & QAR, RPT & Examine the persistence of returns on Bitcoin at different parts on the return distributions & Investors overreact during days of sharp declines in the Bitcoin price and during weeks of market rallies \\ 
\cite{KOCHLING2018} & Efficiency & Bitcoin & Daily & Bit\-coin\-a\-ve\-ra\-ge & LB, RPT, runs test, Bartels, SST, GS, WBAVR, BDS, Hurst exponent & Investigate the effect of futures in market efficiency. & There is no significant switch towards an efficient market \\ 
\cite{ALYAHYAEE2018228} & Efficiency & Bitcoin & Daily & Coindesk & MFDFA & Assess the efficiency of Bitcoin market compared to gold, stock and foreign exchange markets & Bitcoin is more inefficient than the gold, stock and currency markets \\ 
\cite{BOURI2019216} & Efficiency & 14 cryptocurrencies & Daily & Coin\-mar\-ket\-cap & Rolling analysis, CSAD  & Examine the presence of herding behavior & Significant herding behavior varying over time \\ 
\cite{VIDALTOMAS2018} & Efficiency & 65 cryptocurrencies & Daily & Bra\-ve\-New\-Co\-in (BNC)  & CSSD, CSAD & Analyze the existence of herding behavior  & Extreme dispersion of returns explained by rational asset pricing models. Herding during down markets. \\ 
\cite{KAISER2018} & Efficiency & 10 cryptocurrencies & Daily & Coin\-mar\-ket\-cap & Bid-ask spread, GARCH & Test for daily and monthly seasonality in returns, volatility, trading volume and a spread estimator & No consistent and significant calendar effect in returns\\ 
\cite{VIDALTOMAS2018259} & Efficiency & Bitcoin & Daily & Bitstamp, Mt.Gox & AR-CGARCH, AR-CGARCH-M & Examine the semi-strong efficiency of Bitcoin in the Bitstamp and Mt.Gox markets & Bitcoin has no connection to  measures taken by central banks \\ 
\cite{CAPORALE2018} & Efficiency & Bitcoin, Litecoin, Ripple, Dash & Daily & Coin\-mar\-ket\-cap & Independence tests, ANOVA, OLS with dummy variables, trading simulation approach & Examine the day of the week effect & There is no conclusive evidence of inefficiency \\ 
\cite{JIANG2018280} & Efficiency & Bitcoin & Daily & unknown & Rolling window approach & Investigate the time-varying long-term memory in the Bitcoin market & Bitcoin market is inefficient. Returns present strong persistence \\ 
\cite{CHARFEDDINE2019423} & Efficiency & Bitcoin, Ethereum, Litecoin, Ripple & Daily & Coin\-mar\-ket\-cap & LRD (Hurst exponent with various), structural breaks in the returns, splitting sample & Question the true nature of the LRD behavior observed in the returns and volatility & Evidence of LRD in returns and volatility of BTC, LTC and XRP and the volatility of ETH \\ 
\cite{SENSOY201968} & Efficiency & Bitcoin & Intraday (15-minute) & All exchanges & Rolling window approach, permutation entropy & Compare the time-varying weak-form efficiency of Bitcoin prices in US dollars and euro at a high-frequency level & Markets have become more efficient since 2016 \\ 
\cite{CORBET2018} & Efficiency & Bitcoin & Intraday (1-minute) & unknown & EGARCH & Explore as to whether Bitcoin, exhibit similar asymmetric reverting patterns for minutely, hourly, daily and weekly returns & Evidence of several differences in the behavior of Bitcoin price returns according to subperiods and evidence of asymmetric reverting patterns in the Bitcoin price returns \\ 
\cite{Corbet2019} & Literature review & All cryptocurrencies & not applicable & not applicable & Systematic literature review & Provide a systematic review of the empirical literature based on the major topics that have been associated with the market for cryptocurrencies & Finds that there are numerous gaps in the cryptocurrency related literature \\ 
\cite{KOUTMOS201881} & Microstructure & Bitcoin & Daily & Bloomberg & Bivariate VAR & Examine the linkages between Bitcoin returns and transaction activity & Strong linkages between Bitcoin returns  and transaction activity \\ 
\cite{DYHRBERG2018140} & Microstructure & Bitcoin & Intraday (twice a second) & Kraken, Gdax, Gemini & AQS & Examine transactions costs and liquidity of major Bitcoin exchanges & With low spreads and sufficient market depth for average sized transactions, Bitcoin is investible \\ 
\cite{KOUTMOS201897} & Microstructure & Bitcoin & Daily & Bitfinex & ARMA-GARCH, Markov-switching regime & Provide a measure of Bitcoin liquidity uncertainty and to determine market microstructure determinants & Market microstructure variables underlying Bitcoin serve as explanatory variables of Bitcoin liquidity uncertainty \\ 
\cite{KIM2017300} & Microstructure & Bitcoin & Daily & Quandl & Bid-ask spread, multivariate regression & Examine the empirical transaction costs of Bitcoin in international transactions & Transaction cost of Bitcoin is lower than foreign exchange markets \\ 
\cite{ALAOUI2018} & Microstructure & Bitcoin & Daily & Cryp\-to\-com\-pa\-re & Cross-correlation test, MF-DCCA & Study the price-volume cross-correlation  & Price and trading volume mutually interact in a nonlinear way, multifractality is present, Bitcoin market is not efficient \\ 
\cite{HOLUB2019357} & Microstructure & Bitcoin & Daily & Bit\-coin\-charts & Bid-ask spread & Study the global P2P market & Bitcoin bubble's impact on Bitcoin prices in the P2P market is currency and country dependent \\ 
\cite{DELAHORRA201921} & Monetary economics & Bitcoin & Daily & Quandl & Engle-Granger two-step procedure & Analyze the demand for Bitcoin & Bitcoin behaves as a speculative asset in the short term. In the long term, demand might be driven by expectations of Bitcoin's future utility as a medium of exchange \\ 
\cite{Gandal2018} & Monetary economics & Bitcoin& intraday & Bit\-coin\-charts and Mt. Gox & Compare trading volumes in Bit\-coin\-charts and Mt. Gox to verify impact in trading prices& Explore if suspicious trades are linked to movements of bitcoin price & A single trader could excercise significant influence on bitcoin price. Cryptocurrency market is vulnerable to manipulation. \\
\cite{Bohme2015} & Overview &  no applicable & not applicable &not applicable &Overview of cryptocurrency topic &  Discuss bitcoin benefits and costs & Present an overview for a nontechnical audience. Point out risks, regulatory issues, and interactions with the conventional financial system and the real economy. \\
\cite{PLATANAKIS201893} & Portfolio & Bitcoin, Litecoin, Ripple, Dash & Weekly & Coin\-mar\-ket\-cap & MVPO, SR & Examine the diversification benefits of cryptocurrencies & Little difference between na\"ive and optimal diversification \\ 
\cite{SYMITSI2018127} & Portfolio & Bitcoin & Daily & Datastream & VAR conditional mean process, VAR-BEKK-AGARCH, multivariate LB & Study spillover effects between Bitcoin and energy and technology companies & Evidence of unilateral return and volatility spillovers and bidirectional shock influences. Portfolio management implications and benefits. \\ 
\cite{PLATANAKIS2019} & Portfolio & Bitcoin, Litecoin, Ripple, Dash & Weekly & Coin\-mar\-ket\-cap & MVPO, BL(VBCs), SR & Compare different portfolio construction methods using cryptocurrencies & Sophisticated portfolio techniques (Black-Litterman model with VBCs) are preferred when managing cryptocurrency portfolios \\ 
\cite{DYHRBERG201685} & Portfolio & Bitcoin & Daily & Coindesk & GARCH, EGARCH & Explore the financial characteristics of bitcoin using GARCH models & Bitcoin can be classified as something in between gold and the American dollar \\ 
\cite{BAUMOHL2019363} & Portfolio & Bitcoin, Ethereum, Ripple, Litecoin, Stellar Lumens, NEM & Daily & unknown & Quantile cross-spectral approach, standard Pearson's correlations, DMCA & Analyze the connectedness between forex and cryptocurrencies using the quantile & Significant negative dependencies between forex and cryptocurrencies \\ 
\cite{LIU2019200} & Portfolio & 10 cryptocurrencies & Daily & Coin\-mar\-ket\-cap & SR & Examine the investablitiy and role of diversification in cryptocurrency market & Portfolio diversification across different cryptocurrencies can significantly improve investment results \\ 
\cite{BRAUNEIS2019259} & Portfolio & 500 cryptocurrencies & Daily & Coin\-mar\-ket\-cap & MVPO & Assess risk-return benefits of cryptocurrency-portfolios & Combining cryptocurrencies enriches the set of low-risk cryptocurrency investment opportunities \\ 
\cite{JI2019257} & Portfolio & 6 cryptocurrencies & Daily & Coin\-mar\-ket\-cap & VAR, FEVD & Examine connectedness via return and volatility spillovers & Litecoin and Bitcoin are at the centre of the connected network of returns \\ 
\cite{GUESMI2019431} & Portfolio & Bitcoin & Daily & Datastream & DCC-GARCH & Explore the conditional cross effects and volatility spillover between Bitcoin and financial indicators & Bitcoin market allow hedging the risk investment\\ 
\cite{KAJTAZI2019143} & Portfolio & Bitcoin & Daily & Bitcoinity & Mean-CVaR & Explore the effects of adding bitcoin to an optimal portfolio & Bitcoin may help in diversification although it has speculative characteristics \\ 
\cite{URQUHART2017145} & Price clustering & Bitcoin & Daily & Bit\-coin\-charts & Clustering test, conditional effects, standard probit model & Study the price clustering in Bitcoin & There is significant evidence of price clustering at round numbers but there is no significant pattern of returns after the round number. Price and volume have significant positive relationship with price clustering at whole numbers. \\ 
\cite{LI2018} & Price clustering & Bitcoin & Intraday (1-minute) & Bit\-coin\-charts & Chi-squared test, Herfindahl-Hirschman index, OLS & Extend the current literature on price clustering in Bitcoin market & Evidence of clustering for open, high and low prices \\ 
\cite{HU2019337} & Price clustering & Bitcoin, Litecoin, Ripple & Intraday & Bitstamp & Transaction frequency & Investigate intraday price behavior & There is evidence supporting the negotiation and strategic trading hypotheses, but no support for attraction hypothesis \\ 
\cite{AKCORA2018138} & Price discovery & Bitcoin & Daily & Coinbase & HFG, GARCH  & Model the network with a high fidelity graph to characterize the flow of information & Identification of certain sub-graphs with predictive influence on Bitcoin price and volatility\\ 
\cite{BRAUNEIS201858} & Price discovery & 73 cryptocurrencies & Daily & Coin\-mar\-ket\-cap & KS, GARCH, (LB, VR, BDS, Hurst exponent), MOE, TR & Investigate efficiency / predictability  and to asses liquidity of cryptocurrencies & Efficiency is positively related to liquidity \\ 
\cite{URQUHART201840} & Price discovery & Bitcoin & Intraday (5-minute) & Bit\-coin\-charts & RV, VAR & Study the attention of Bitcoin by employing Google Trends data & Attention of Bitcoin is influenced by the previous day's high realized volatility and volume \\ 
\cite{KAPAR201962} & Price discovery & Bitcoin & Daily & Coindesk & IS, CS & Analyze the Bitcoin price discovery process & The Bitcoin futures market dominates the price discovery process \\ 
\cite{SHEN2019118} & Price discovery & Bitcoin & Intraday (5-minute) & Bit\-coin\-charts & VAR, Granger causality test & Examine the link between investor attention and Bitcoin returns, trading volume and realized volatility & The number of tweets is a significant driver of next day trading volume and realized volatility \\ 
\cite{SUN2018} & Price discovery & 42 cryptocurrencies & Daily & Investing & LightGBM (GBDT), SVM, RF & Forecast the price trend  & LightGBM algorithm outperforms other methods \\ 
\cite{TROSTER2018} & Price discovery & Bitcoin & Daily & Coindesk & GARCH, GAS, VaR & Model and forecast bitcoin returns and risk & Heavy-tailed GAS models improve goodness-of-fit and forecast performance of bitcoin returns and risk \\ 
\cite{DEMIR2018145} & Price discovery & Bitcoin & Daily & Coindesk & VAR, OLS  & Analyze the prediction power of the economic policy uncertainty (EPU) index on the daily Bitcoin returns & EPU has a predictive power on Bitcoin returns, serving as a hedging tool against uncertainty \\ 
\cite{FENG201863} & Price discovery & Bitcoin & Daily & Bit\-coin\-charts & OSI & Propose a novel indicator to assess informed trades ahead of cryptocurrency-related events & Evidence of informed trading in the Bitcoin market prior to both positive and negative large events \\ 
\cite{PANAGIOTIDIS2018235} & Price discovery & Bitcoin & Daily & Coindesk & LASSO & Examine the significance of twenty-one potential drivers of bitcoin returns & Search intensity and gold returns are the most important variables for bitcoin returns \\ 
\cite{BOURI2019340} & Price discovery & Bitcoin, Ripple, Ethereum, Litecoin, Nem, Dash, Stellar & Daily & Coin\-mar\-ket\-cap & Granger causality & Extend the understanding on the Granger causality from trading volume to the returns and volatility & Evidence of Granger causality from trading volume to the returns \\ 
\cite{AALBORG2019255} & Price discovery & Bitcoin & Intraday (10-minute) & Bit\-coin\-charts & Heterogeneous AR, HAR-RV & Study which variables can explain and predict the return, volatility and trading volume of Bitcoin & Trading volume can be predicted from Google searches, but none of the considered variables can predict returns \\ 
\cite{DASTGIR2019160} & Price discovery & Bitcoin & Weekly & Investing & Granger Causality & Examines the causal relationship between Bitcoin attention (measured by the Google Trends search queries) and Bitcoin returns & Bidirectional causality mainly exists in both tails \\ 
\cite{PANAGIOTIDIS2019220} & Price discovery & Bitcoin & Daily & Coindesk & VAR, FAVAR, PCA & Examine the effects of shocks on bitcoin returns & Evidence of a significant interaction between bitcoin and traditional stock markets, weak interaction with FX markets and the macroeconomy \\ 
\cite{BLEHER2019147} & Price discovery & 12 cryptocurrencies & Intraday (hourly) & Cryp\-to\-com\-pa\-re & VAR,  Granger-causality & Evaluate the usefulness of Google search volume to predict returns and volatility of multiple cryptocurrencies & Returns are not predictable, volatility is partly predictable \\ 
\cite{DYHRBERG2016139} & Safe-haven & Bitcoin & Daily & Coindesk & Asymmetric GARCH & Explore the hedging capabilities of bitcoin & Bitcoin possess some of the same hedging abilities as gold \\ 
\cite{BOURI201787} & Safe-haven & Bitcoin & Daily & Coindesk & OLS, Wavelet decomposition & Examine whether Bitcoin can hedge global uncertainty & Bitcoin does act as a hedge against uncertainty in the short horizon \\ 
\cite{BOURI2017192} & Safe-haven & Bitcoin & Daily & Thomson Reuters & DCC & Examine whether Bitcoin can act as a hedge and safe haven for major world stock indices, bonds, oil, gold, the general commodity index and the US dollar index & Bitcoin is a poor hedge and is suitable for diversification purposes only \\ 
\cite{Smales2018} & Safe-haven & Bitcoin & Daily & Da\-ta.\-bit\-co\-ini\-ty, Block\-cha\-in.\-com & Correlation with other assets  & Study whether Bitcoin characteristics in a period of relative calm (2011-2017) is coherent with a safe-haven asset & Bitcoin is not currently a safe haven, although its low correlation with traditional assets \\ 
\cite{BAUR2018103} & Safe-haven & Bitcoin & Daily & Coindesk & GARCH, EGARCH, TGARCH & Analyze the relationship between Bitcoin, gold and the US dollar & Bitcoin exhibits distinctively different return, volatility and correlation characteristics compared to other assets \\ 
\cite{KLEIN2018105} & Safe-haven & Bitcoin & Daily & Coindesk & APARCH, FIAPARCH, BEKK-GARCH & Compares Gold and Bitcoin from an econometric perspective  & Bitcoin and Gold feature fundamentally different properties as assets and linkages to equity markets \\ 
\cite{URQUHART201949} & Safe-haven & Bitcoin & Intraday (hourly) & Bit\-coin\-charts & DCC, ADCC, GARCH, GJRGARCH, EGARCH & Investigate whether Bitcoin can act as a hedge or safe haven against world currencies & Bitcoin can be considered as hedge and diversifier for currency investors \\ 
\cite{KATSIAMPA20173} & Volatility & Bitcoin & Daily & Coindesk & AR, EGARCH, TGARCH, APARCH, CGARCH, ACGARCH & Study the ability of several GARCH models to explain Bitcoin price volatility & The optimal model in terms of goodness-of-fit to the data is the AR-CGARCH \\ 
\cite{BAUR2018148} & Volatility & 20 cryptocurrencies & Daily & Coin\-mar\-ket\-cap & TGARCH, AR, QAR & Analyze asymmetric volatility effects for the 20 largest cryptocurrencies & Volatility increases more in response to positive shocks than to negative shocks \\ 
\cite{CORBET201823} & Volatility & Bitcoin & Intraday (1-minute) & Thomson Reuters & Mood statistic, Lepage statistic, OLS, IS, CS, ILS & Investigate the effect of the introduction of Bitcoin futures & The introduction of Bitcoin futures has increased the spot volatility of Bitcoin \\ 
\cite{CHAIM2018158} & Volatility & Bitcoin & Daily & unknown & SV, \cite{QuPerron2013} and \cite{laurini2016multivariate} & Estimate stochastic volatility models with jumps to volatility and returns & Jumps to volatility are permanent, jumps to returns are contemporaneous, volatility was highest in late 2013 and during 2014, big jumps to mean returns are negative and related to hacks and forks \\ 
\cite{KHUNTIA2018} & Volatility & Bitcoin & Hourly & Bit\-coin\-charts & MFDFA &  Evaluate the adaptive pattern of long memory in the volatility of intra-day bitcoin returns and to test the impact of the trading volume on time-varying long memory & Long memory exists and fluctuates over time, the time-varying pattern of long memory is coherent with AMH \\ 
\cite{PHILLIP201995} & Volatility & 149 cryptocurrencies & Daily & Brave New Coin (BNC)  & JBAR-SV-GLR & Study some stylized facts about the variance measures of Cryptocurrencies & Volatility of Cryptocurrencies can be  measured with fast moving autocorrelation functions, as opposed to smoothly decaying functions for fiat currencies \\ 
\cite{TAN2018} & Volatility & 102 cryptocurrencies & Daily & Coin\-mar\-ket\-cap & GK, ABL-CARR & Measure and model volatilities & There is evidence of volatility persistence and leverage effects improving predictability of volatility, reducing risk and diminishing the level of speculation in cryptocurrency market \\ 
\cite{ARDIA2019266} & Volatility & Bitcoin & Daily & Datastream & MSGARCH, VaR & Test the presence of regime changes in the GARCH volatility dynamics  & Daily log-returns exhibit regime changes in their volatility dynamics \\ 
\cite{MENSI2019222} & Volatility & Bitcoin, Ethereum & Daily & Coindesk & GARCH, FIGARCH, FIAPARCH, HYGARCH, Markov-switching dynamic regression & Explore the impacts of structural breaks on the dual long memory levels of Bitcoin and Ethereum price returns & Evidence of dual long memory property of Bitcoin and Ethereum \\ 
\cite{Katsiampa2018} & Volatility & Bitcoin, Ethereum & Daily & Coin\-mar\-ket\-cap & Bivariate Diagonal BEKK & Investigate the volatility dynamics of the two major cryptocurrencies & Evidence of interdependency in the cryptocurrency market. Conditional volatility and correlation are responsive to major news \\ 
\cite{YI201898} & Volatility & 52 cryptocurrencies & Daily & Coin\-mar\-ket\-cap & Volatility spillover index (GVD), LASSO-VAR & Examine both static and dynamic volatility connectedness  & Connectedness fluctuates cyclically and has shown a rise trend since the end of 2016 \\ 
\cite{FANG201929} & Volatility & Bitcoin & Daily & Coindesk & GARCH-MIDAS, DCC-MIDAS & Assess whether the long-run volatilities of Bitcoin, global equities, commodities, and bonds are affected by global economic policy uncertainty & The long-term volatility of Bitcoin, equities, and commodities are significantly affected by economic policy uncertainty, although the effect on the volatility of Bitcoin is different from the other assets \\ 
\cite{GILLAIZEAU201986} & Volatility & Bitcoin & Daily & Bit\-coin\-charts & GVD & Identify and characterize the givers and the receivers of volatility in crossmarket Bitcoin prices and to discuss diversification strategies & Bitcoin prices depict strong dynamic spillover in volatility, especially during episodes of high uncertainty \\ 
\label{tab:detailPapers}
\end{longtable}
\end{landscape}

\newgeometry{top=1in,bottom=1in,right=0.75in,left=0.75in}

\small
\begin{table}[htbp]
  \centering
  \caption{List of acronyms used in Table \ref{tab:detailPapers}}
  \begin{tabular}{ll}
		\toprule
    Acronym & Name \\
		\midrule
    ABL-CARR  &  Asymmetric bilinear Conditional Autoregressive Range  \\
    ACGARCH  &  Asymmetric Component GARCH  \\
    BL(VBCs)  &  Black–Litterman portfolio optimization with variance-based constraints  \\
    CGARCH  &  Component GARCH \\
    CSAD  &  Cross-sectional absolute standard deviations  \\
    CSSD  &  Cross-sectional standard deviation of returns \\
    DCC   & Dynamic conditional correlation  \\
    DFA   &  Detrended Fluctuation Analysis \\
    DMCA  &  Detrended moving-average cross-correlation analysis \\
    ELW   &  Exact local Whittle \\
    FCVAR  &  Fractionally cointegrated VAR \\
    FIAPARCH  &  Fractionally integrated asymmetric power ARCH \\
    FIGARCH  &  Fractionally integrated GARCH \\
    GAS   &  Generalized Auto-regressive Score \\
    GK    & Garman and Klass volatility measures  \\
    GLR   &  Gegenbauer Log Range \\
    HYGARCH  &  Hyperbolic GARCH \\
    JBAR  &  Jump Buffered Autoregressive model \\
    LASSO  &  Least Absolute Shrinkage and Selection Operator \\
    LB    &  Ljung–Box test \\
    LightGBM  &   Light Gradient Boosting Machine \\
    LRD   & Long Range Dependence \\
    MF-DCCA  &  Multifractal detrended cross-correlations analysis \\
    MSGARCH  &  Markov–switching GARCH  \\
    MVPO  &  Mean–variance portfolio optimization \\
    RPT   &  robustified portmanteau test \\
    SR    & Sharpe ratio \\
    VAR   &  Vector autoregression \\
    VaR   &  Value at risk test \\
    VAR-BEKK-AGARCH  &  asymmetric BEKK Generalized Autoregressive Conditional Heteroskedasticity \\
    VR    & Variance Ratio Test \\
    WBAVR  & Wild bootstrapped automatic VR test \\
		\bottomrule
    \end{tabular}%
  \label{tab:acronyms}%
\end{table}%

\end{document}